# The light carrying orbital angular momentum through time-varying scattering media using dual orthogonal-polarization channels

**Heshen Li,[1] Jin Wei,[1] Tianshun Zhang,[1] and Wen Chen[1,2,a)]**

**AFFILIATIONS**

[1]Department of Electrical and Electronic Engineering, The Hong Kong Polytechnic University, Hong Kong, China
[2]Photonics Research Institute, The Hong Kong Polytechnic University, Hong Kong, China
[a)]**Author to whom correspondence should be addressed:** owen.chen@polyu.edu.hk

## ABSTRACT

Orbital angular momentum (OAM) of light can be used as a degree of freedom for optical communication. However, the reliable transmission of OAM beams through time-varying scattering media remains challenging. In this paper, we report an approach for OAM transmission through time-varying scattering media using dual orthogonal-polarization channels. A perfect vortex beam (PVB) is generated and transmitted at one channel, and the plane wave is employed as a reference beam at another channel. With the cross-correlation obtained between the recorded speckle patterns at each single-shot measurement, the data can be retrieved without *prior* knowledge about time-varying scattering media. It is revealed that the multiplexed PVB can be applied to generate recognizable coherent-superposition patterns in cross-correlation images. Experimental results demonstrate that the proposed method can be applied to transmit the data through time-varying scattering media, and high robustness can be achieved. The proposed method could open up an avenue for the development of OAM transmission in harsh environments.

## I. INTRODUCTION

Light carrying orbital angular momentum (OAM) has attracted much current attention, and can be used to enhance the capacity of optical communications.[1,2] In theory, OAM beams can carry infinite orthogonal eigenstates, showing potential for high-capacity optical communications.[3,4] However, it is inevitable that OAM beams would be distorted in free space, e.g., time-varying scattering media.[5]

Existing methods for light transmission in scattering media can be categorized as active wavefront modulation[6–9] and passive exploitation of the scattered light field.[5,10,11] The latter approaches could include transmission matrix,[10] vortex memory effect,[12] and cross-correlation,[13] etc. For instance, with the transmission matrix, the scattered optical field can be retrieved by performing inverse transformation to implement OAM demultiplexing.[10,14–16] Using vortex memory effect, reference speckle patterns can be pre-recorded according to different OAM beams.[12] However, these approaches usually require calibration of the scattering media or multiple reference measurements, and their feasibility and applicability could be limited especially in time-varying scattering media. Alternatively, cross-correlation of the scattered speckle patterns can be calculated to retrieve data.[13,17,18] It originated from the inherently statistical properties of speckle fields,[18–21] and optical information is retrieved from speckle patterns.[22,23] Vanitha et al.[13] demonstrated that the cross-correlation image with a ring-like pattern can be calculated between two speckle patterns corresponding to two perfect vortex beams (PVBs) with different topological charges (TCs). Ma et al.[17] realized OAM mode sorting and transmission. However, an axial offset of the receiver would induce lateral scaling effect in the cross-correlation images, severely limiting reliable transmission over varying propagation distances. Zhang et al.[24] reported dual-TC multiplexing for OAM-encoded transmission. This method needs much effort to create a dataset. In addition, most studies focus on beam transmission under simple conditions (e.g., static scattering media), restricting their wide deployments in real-world scenarios. In recent years, vectorized optical fields were extensively studied for optical data transmission,[25–29] and mutually orthogonal polarization channels[26,30–32] were designed and applied. However, the vortex beam with a single TC and a fixed position for the detector are usually required.[31]

In this paper, we report OAM transmission through time-varying scattering media using dual orthogonal-polarization channels. A multiplexed PVB and the plane wave are generated in optical channels to serve as the signal and reference beams, respectively. It is experimentally observed that the mutual coherence function (MCF) can preserve the vortex phase, and the controllable coherent superposition can be achieved. The MCF coherent superposition is observed in experiments via calculating cross-correlation between the recorded speckle patterns at each single-shot measurement using a division-of-focal-plane (DoFP) polarization camera, and then the data can be decoded from cross-correlation images. Experimental results demonstrate that 6-bit data can be encoded using PVB with 4 TCs, and the proposed method can achieve high-fidelity OAM transmission without *prior* knowledge about time-varying scattering media. It is also illustrated that the proposed method is robust against the misalignments of the detector. This work offers a promising and robust way for OAM transmission in harsh environments, and could have broad application prospects.

## II. METHOD

### A. Vortex phase in MCF

The propagation of an OAM beam can be described by wave diffraction, and the scattering effect could be inevitable in free space. Nevertheless, the information carried by OAM beams would not be completely destroyed even under strong scattering conditions. The scattering effect can be regarded as an unknown converter that projects OAM space onto a measurable light field,[12] and different algorithms can be developed to retrieve the OAM carried by the light beam. In this study, the designed procedure of PVB transmission through time-varying scattering media is shown in Fig. 1. The P-polarized multiplexed PVB is used as the signal beam, and the S-polarized plane wave is used as the reference beam. Then, the signal and reference beams are combined to illuminate a rotating ground glass (RGG). The speckle patterns corresponding to the PVB and the plane wave are recorded by using a DoFP polarization camera at a single-shot measurement. At the receiver, a cross-correlation image between two recorded speckle patterns can be obtained. By extracting the features from the cross-correlation image, the data encoded in the multiplexed PVB can be retrieved.

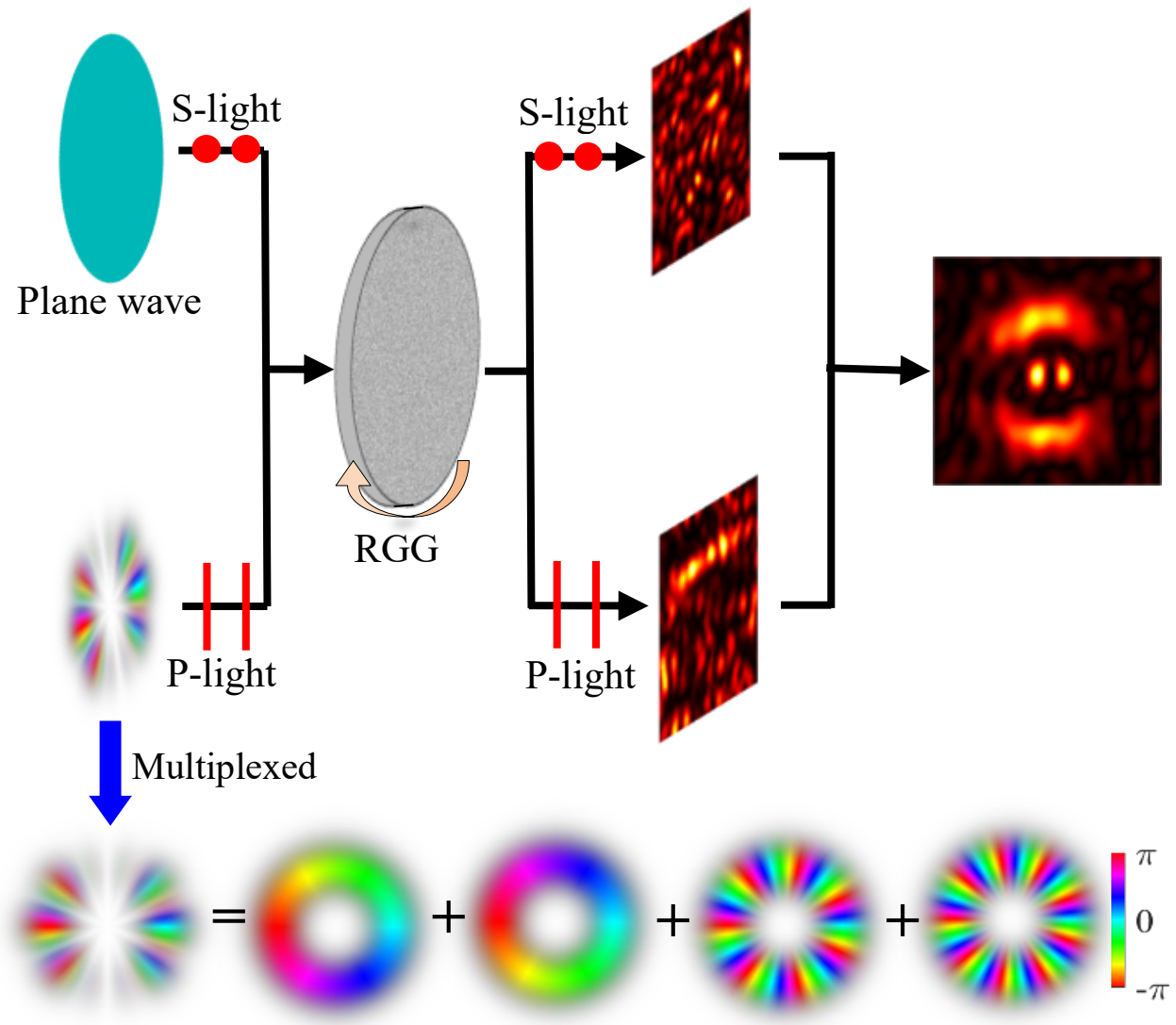

**FIG. 1.** A schematic of data encoding and the retrieval of a cross-correlation image using the proposed method. Here, the multiplexed PVB with 4 TCs ($|{-1}\rangle+|1\rangle+|7\rangle+|9\rangle$) is given as an example. RGG: rotating ground glass.

The PVB with a Gaussian envelope can be described by[33,34]

$$u_\ell(\mathbf{r}) = \exp\left[-\frac{(r-r_0)^2}{\Delta r^2}\right]\exp(i\ell\varphi) = u_0(\mathbf{r})\exp(i\ell\varphi), \tag{1}$$

where $i=\sqrt{-1}$, $\ell$ denotes the TC, $r_0$ denotes the radius of the PVB annulus, $\Delta r$ denotes the width of the annulus, and $\mathbf{r}$ denotes a position vector consisting of the radius $r$ and the azimuth $\varphi$.

After wave propagation through time-varying scattering media as shown in Figs. 1 and 2(a), four speckle patterns, i.e., four sub-images, are recorded at each single-shot measurement by using the DoFP polarization camera. The MCF obtained between the two speckle patterns corresponding to the signal and reference beams (see Supplementary Note S1 for details) can be described by

$$\Gamma_\ell(\Delta\mathbf{s},z) = C\exp\left[\frac{ik}{2z}\left(s_1^2-s_2^2\right)\right]U_\ell\left(\frac{k\Delta\mathbf{s}}{z}\right), \tag{2}$$

where $C=\lambda^{-2}z^{-2}$, $\lambda$ denotes the wavelength, $z$ denotes the axial distance between the RGG and camera, $k$ denotes the wavenumber, $\Delta\mathbf{s}=\mathbf{s}_1-\mathbf{s}_2$, $\mathbf{s}_1$ and $\mathbf{s}_2$ denote the position vectors on the sub-image planes respectively corresponding to the multiplexed PVB and the plane wave, and $U_\ell$ denotes the Fourier spectrum of $u_\ell$. The Fourier spectrum $U_\ell$ can be described by

$$\begin{aligned}
U_\ell\left(\frac{k\Delta\mathbf{s}}{z}\right) &= \iint u_0(\mathbf{r})\exp(i\ell\varphi)\exp\left[-\frac{ik}{z}r\Delta s\cos(\theta_0-\varphi)\right]rdrd\varphi \\
&= \int_0^\infty\int_0^{2\pi} u_0(\mathbf{r})\exp\left[i\ell(\theta_0-\xi)\right]\exp\left(-\frac{ik}{z}r\Delta s\cos\xi\right)rdrd\xi \\
&= 2\pi i^{-\ell}\exp(i\ell\theta_0)\int_0^\infty u_0(r)J_\ell\left(\frac{k}{z}r\Delta s\right)rdr \\
&= 2\pi i^{-\ell}\exp(i\ell\theta_0)K_\ell\left(\frac{k\Delta s}{z}\right),
\end{aligned} \tag{3}$$

where $\Delta s$ and $\theta_0$ respectively denote radial and azimuthal coordinates of $\Delta\mathbf{s}$, $\xi=\theta_0-\varphi$, $J_\ell(\cdot)$ denotes the $\ell$-th order Bessel function of the first kind, and $K_\ell$ is given by the $\ell$-th order Hankel transform of $u_0(r)$. The term $\exp(i\ell\theta_0)$ exists in Eq. (3), verifying that the MCF contains vortex phase. In addition, the radial and azimuthal coordinates are effectively separated corresponding to radial amplitude and vortex phase, respectively.

### B. MCF coherent superposition

Since the Fourier and Hankel transforms are linear, the MCF of a signal beam consisting of multiple PVBs is a linear superposition of the MCFs corresponding to each PVB. The multiplexed PVB $h(\mathbf{r})$ containing $N$ TCs can be described by

$$h(\mathbf{r}) = \sum_{n=1}^{N} a_n u_0(\mathbf{r})\exp(i\ell_n\varphi), \tag{4}$$

where $a_n$ denotes complex-valued coefficients.

Then, its corresponding MCF can be obtained by

$$\Gamma_{total} = \sum_{n=1}^{N} a_n \Gamma_{\ell_n}, \tag{5}$$

The complex Gaussian moment theorem[17,22,23] is utilized to describe the relationship between fourth-order cross-correlation[35,36] and the MCF (see Supplementary Note S1 for details), which can be described by

$$\left\langle I_\ell(\mathbf{s})I_{plane}(\mathbf{s}+\Delta\mathbf{s})\right\rangle - \left\langle I_\ell(\mathbf{s})\right\rangle\left\langle I_{plane}(\mathbf{s}+\Delta\mathbf{s})\right\rangle = \left|\Gamma_\ell(\Delta\mathbf{s})\right|^2, \tag{6}$$

where $\langle\,\rangle$ denotes ensemble averaging, $*$ denotes complex conjugate, and $I_\ell$ and $I_{plane}$ denote the recorded speckle patterns at each single-shot measurement corresponding to the PVB and the plane wave, respectively.

For the PVB carrying a single TC, a ring-like pattern in the cross-correlation image is obtained, and its radius is determined by $|\ell|$, $u_0$, $k$ and $z$. Here, the vortex phase cannot be directly observed, since the measured intensity is proportional to the squared amplitude of MCF. When the multiplexed PVB carrying multiple TCs is used, coherent superposition of individual MCFs becomes possible as indicated in Eq. (5), leading to petal-like patterns in the cross-correlation image. When the MCF is generated by using the multiplexed PVB with close TCs ($|\ell|$), the coherent superposition can be clearly observed. For instance, the TCs $|{-1}\rangle$ and $|1\rangle$ can be used in Fig. 1 to render dual-petal coherent superposition with a small radius, while the usage of TCs $|7\rangle$ and $|9\rangle$ leads to dual-petal coherent superposition with a large radius. Moreover, since complex-valued coefficients $a_n$ are used in Eq. (4), different phase values can be assigned to $a_n$, offering the generation of an additional degree of freedom to control the azimuthal distribution of the petals in the cross-correlation image.

## III. EXPERIMENTAL RESULTS AND DISCUSSION

An optical setup is designed to validate the proposed method as shown in Fig. 2(a). A 532.0-nm laser (MGL-III-532, 200 mW) is expanded and collimated. A half-wave plate (HP) is used to adjust the polarization of the light beam for illuminating a spatial light modulator (SLM, FSLM-2K70-P02HR, CAS Microstar). The SLM is loaded with a phase-only hologram generated by using the Jacobi-Anger identity.[37–40] Here, only the P-light is modulated by the SLM, and the S-light remains unaltered. Therefore, the first-order diffraction of the P-polarized beam can carry the target light field, and the S-polarized plane wave is used as the reference beam. Lenses 1 and 2 with a focal length of 200.0 mm are used to form a 4f system, and a Mach-Zehnder interferometer is applied to combine the P-polarized signal beam and S-polarized reference beam. Then, the light beam illuminates the RGG (600 grit, a rotating speed of 0.25 rpm) used to emulate time-varying scattering media. A DoFP polarization camera (MER2-503-36U3M POL, Daheng Image Vision) as shown in Fig. 2(b) is placed 300.0 mm behind the RGG to record four speckle patterns at each single-shot measurement. Here, the RGG induces negligible polarization effects (see Supplementary Note S2 for details). The detector is covered with micro-polarizers at four polarization angles (i.e., 0°, 45°, 90° and 135°) arranged into 2×2 super-pixel units. Four sub-images (speckle patterns) corresponding to the polarization angles of 0°, 45°, 90° and 135° can be extracted from each measurement. The sub-images recorded with polarization angles of 0° and 90° respectively correspond to the S- and P-polarized light, and the sub-images with polarization angles of 45° and 135° contain interference between the S- and P-polarized light beams.

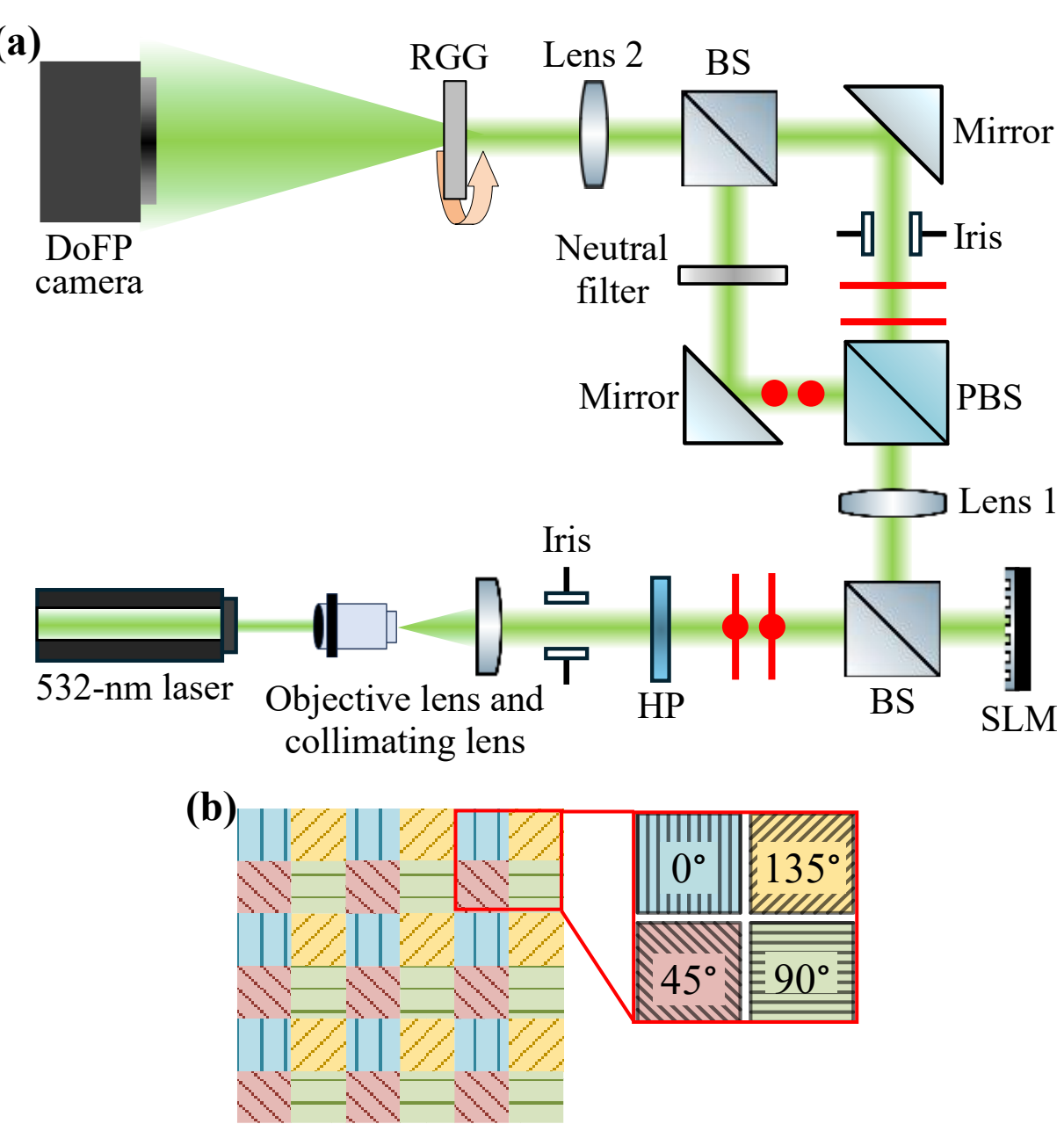

**FIG. 2.** (a) A schematic of the designed optical setup. HP: half-wave plate; BS: beam splitter; PBS: polarization beam splitter. (b) A schematic of the DoFP polarization camera.

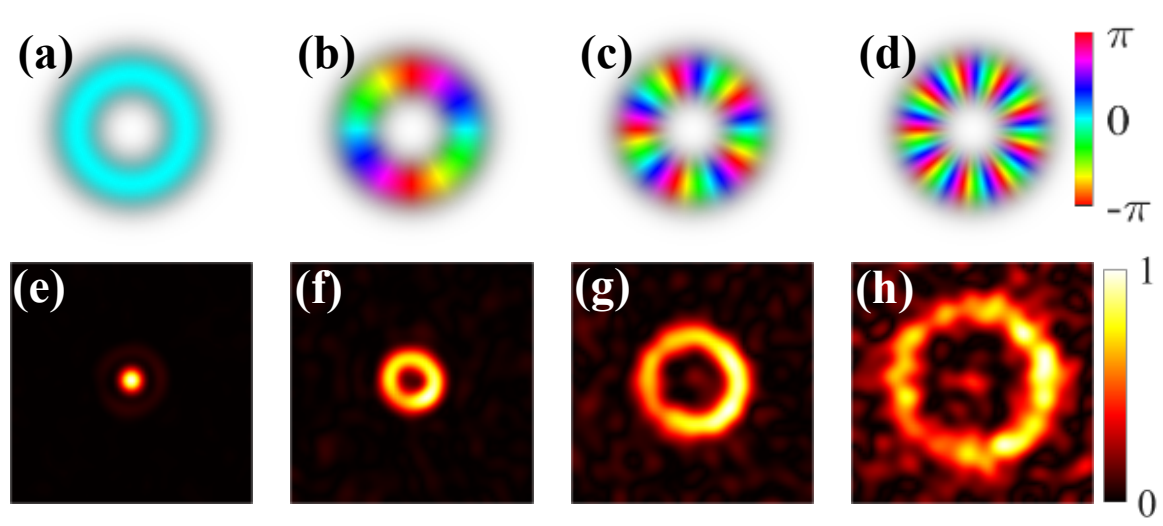

**FIG. 3.** Experimental results for the transmission of single-mode PVB through time-varying scattering media. (a)-(d) The single-mode PVB with TCs of $|0\rangle$, $|2\rangle$, $|5\rangle$, $|9\rangle$, respectively. (e)-(h) The retrieved cross-correlation images respectively corresponding to (a)-(d).

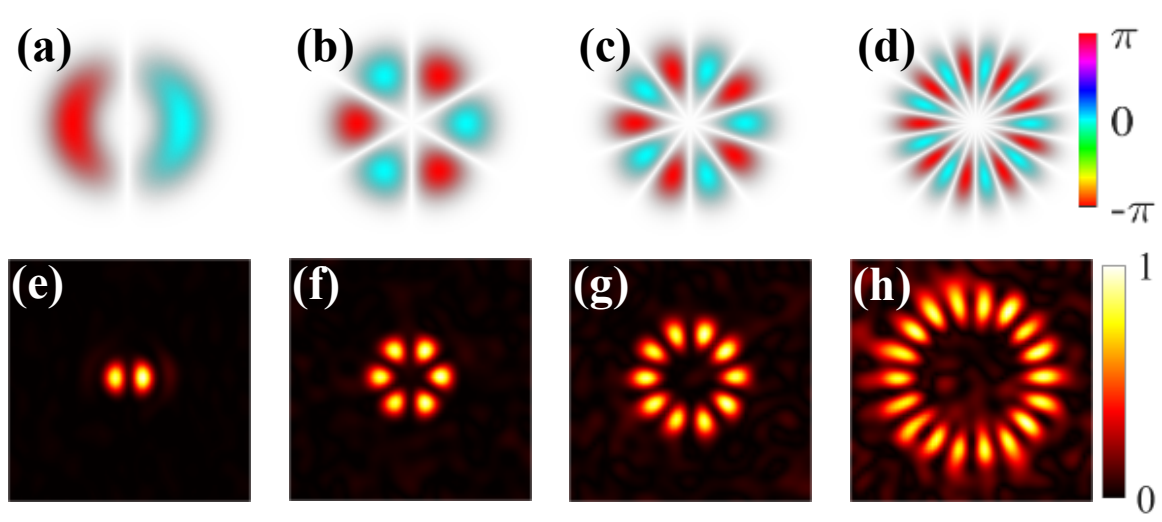

**FIG. 4.** Experimental results for the transmission of the multiplexed PVB with 2 TCs through time-varying scattering media. (a)-(d) The multiplexed PVBs with TCs of $|-1\rangle+|1\rangle$, $|-3\rangle+|3\rangle$, $|-5\rangle+|5\rangle$ and $|-9\rangle+|9\rangle$, respectively. (e)-(h) The retrieved cross-correlation images respectively corresponding to (a)-(d).

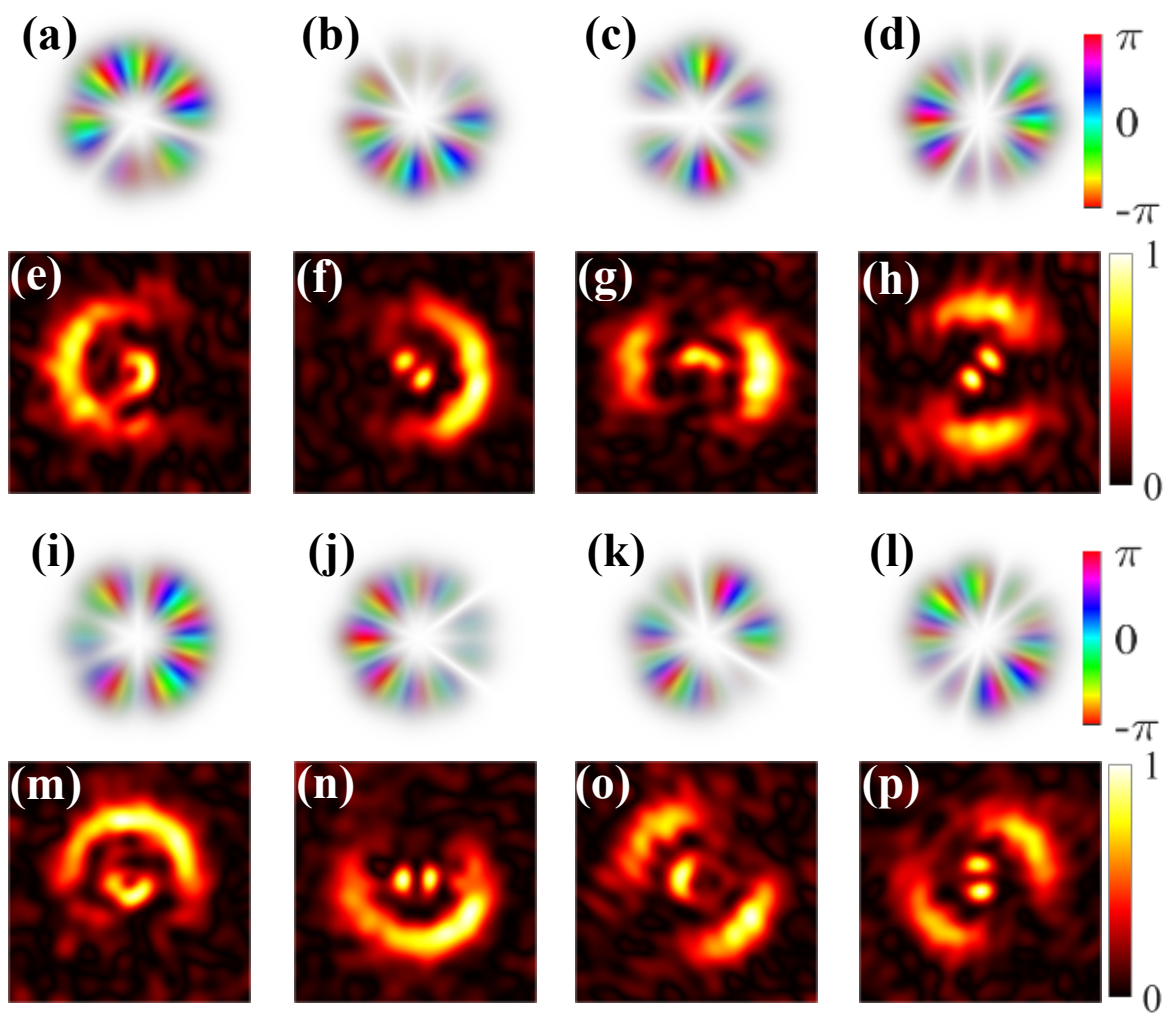

**FIG. 5.** Experimental results for the transmission of the multiplexed PVB with multiple TCs through time-varying scattering media. Here, phase values $(0, \pi/2, \pi, 3\pi/2)$ and the weights (see Supplementary Note S3 for details) are used to generate complex-valued coefficients $a_n$ in Eq. (4). (a) $|1\rangle+i|2\rangle+|7\rangle-i|8\rangle$, (b) $|-1\rangle+i|1\rangle+|7\rangle+i|8\rangle$, (c) $|1\rangle+|2\rangle+|7\rangle-|9\rangle$, (d) $|-1\rangle-i|1\rangle+|7\rangle+|9\rangle$, (i) $|1\rangle-|2\rangle+|7\rangle+|8\rangle$, (j) $|-1\rangle+|1\rangle+|7\rangle-|8\rangle$, (k) $|1\rangle-i|2\rangle+|7\rangle-i|9\rangle$, (l) $|-1\rangle-|1\rangle+|7\rangle+i|9\rangle$. (e)-(h) and (m)-(p) The retrieved cross-correlation images respectively corresponding to (a)-(d) and (i)-(l).

Figure 3 shows experimental results for the transmission of single-mode PVB through time-varying scattering media. In experiments, the PVB has a radius of $r_0 = 1.8$ mm and a Gaussian envelope width of $\Delta r = 0.83$ mm. As shown in Figs. 3(e)–3(h), the ring radius in the cross-correlation images increases with TC. The experimental results align with the descriptions in Eq. (3). As given in Eq. (6), a cross-correlation image is determined by the amplitude of the MCF, and the vortex phase cannot be directly observed. Hence, the cross-correlation images in Fig. 3 render ring-like patterns without the azimuth information. To reveal that MCF contains vortex phase, two PVBs with opposite TCs are multiplexed to show coherent superposition. Figure 4 shows experimental results for the transmission of the multiplexed PVB through time-varying scattering media. With the MCF coherent superposition, the cross-correlation images exhibit petal-like patterns. The total number of petals at a layer is equivalent to the absolute value of the TC difference. It is revealed that the vortex phase in MCF is involved in controlling the azimuthal intensity distribution in the cross-correlation image. It can also be seen in Fig. 4 that the ring radius in the retrieved cross-correlation images increases with TC, originating from the main peak positions of Bessel functions of different orders in Eq. (3).

The proposed method is also applied to form dual-layer cross-correlation images using the multiplexed PVB with 4 TCs, as shown in Fig. 5. The MCFs corresponding to the multiplexed PVB with close TCs (absolute values) would undergo coherent superposition. In contrast, coherent superposition does not clearly exist when the multiplexed PVB with substantially different TCs is used in the optical setup (see Supplementary Note S4 for details). Here, the complex-valued coefficients $a_n$ are used as described in Eq. (4), and different phase values are assigned to $a_n$ to control the azimuthal intensity distribution (see Supplementary Note S5 for details).

The number and azimuthal position of petals in each layer of the cross-correlation image can be extracted via Fourier analysis of the azimuthal intensity distributions using four-step phase-shifting.[41] Assuming that $f(\theta_0)$ represents the azimuthal intensity distribution of a layer in the retrieved cross-correlation image (see Supplementary Note S6 for details), Fourier coefficients $B_m$ can be calculated by

$$D_\beta^{(m)} = \int_0^{2\pi} f(\theta_0)\cos(m\theta_0+\beta)d\theta_0, \tag{7}$$

$$B_m = D_0^{(m)} - D_\pi^{(m)} + i\left(D_{\pi/2}^{(m)} - D_{3\pi/2}^{(m)}\right), \tag{8}$$

where $m$=1,2,3,…, and $\beta = 0,\ \pi/2,\ \pi,\ 3\pi/2$. For instance, in Fig. 5, by comparing coefficients $|B_m|$, $m$ is determined to be 1 when $|B_1|$ is the maximum value. Similarly, $m$ is determined to be 2 when $|B_2|$ is the maximum value. After the total number of petals $m$ is determined, the rotational angle of a layer in the retrieved cross-correlation image (e.g., $0,\ \pi/2,\ \pi,\ 3\pi/2$ in Fig. 5) can be obtained using the phase part of the determined coefficient $B_m$. At each layer of the cross-correlation images in Figs. 5(e)–5(h) and 5(m)–5(p), 1-bit data can be decoded using the determined value $m$, and 2-bit data can be decoded using the rotational angle. Hence, 6-bit data are decoded correspondingly from the retrieved cross-correlation image.

Figure 6 shows the retrieval of a 6-bit image using the proposed method. The 6-bit image with 64×64 pixels as shown in Fig. 6(a) is encoded into 4096 multiplexed PVBs to be transmitted through time-varying scattering media in Fig. 2(a). Figure 6(b) shows the retrieved image obtained at the receiver. Here, all 4096 6-bit data are decoded accurately. Figure 6(c) shows the phase values retrieved from the azimuthal intensity distributions obtained based on Eqs. (7) and (8). It can be seen in Fig. 6(c) that the phase terms are effectively separated, and data decoding can be reliably implemented.

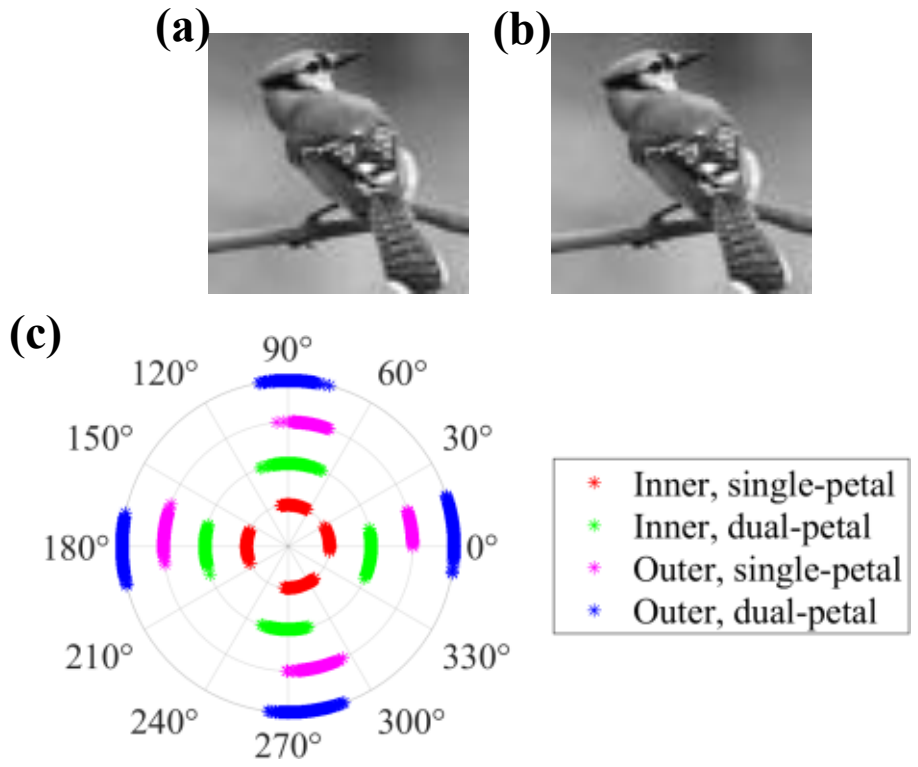

**FIG. 6.** The retrieval of a 2D image in time-varying scattering media using the proposed method. (a) The original image. (b) The decoded image. (c) The phase values retrieved from the azimuthal intensity distributions at each layer of the cross-correlation images.

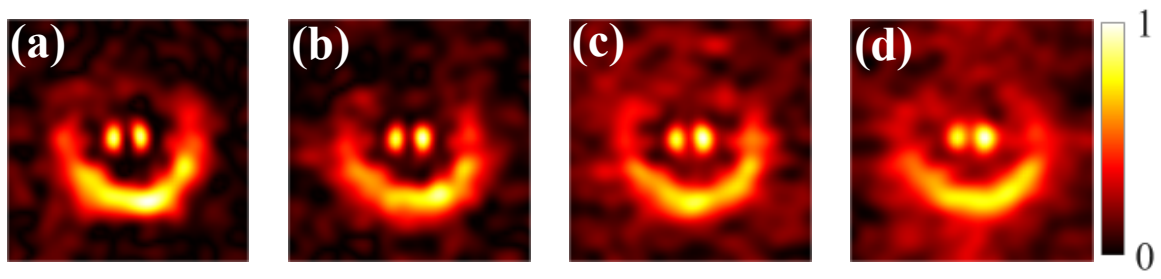

**FIG. 7.** Experimental results for the analysis of lateral misalignments of the detector. The retrieved cross-correlation images for the multiplexed PVB with TCs of $|-1\rangle+|1\rangle+|7\rangle-|8\rangle$ when there is a lateral misalignment of (a) 0, (b) 10.0, (c) 30.0 and (d) 50.0 mm.

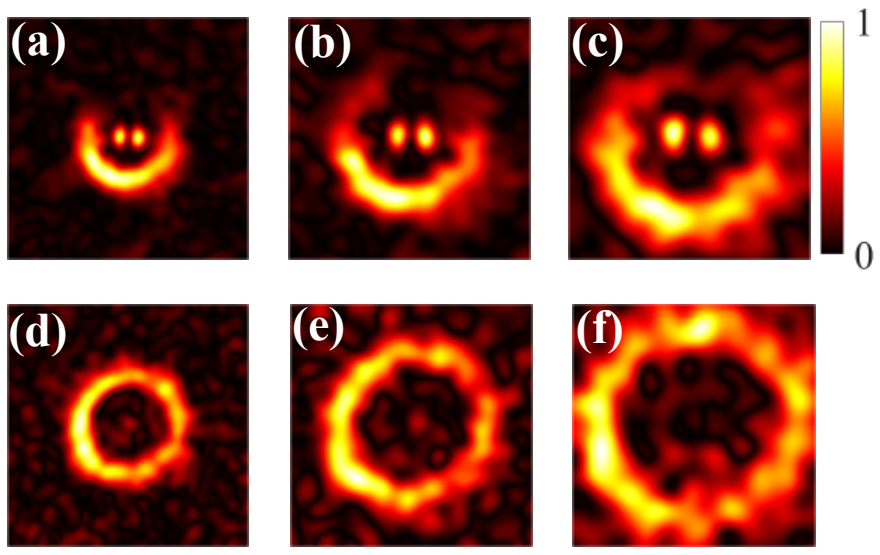

**FIG. 8.** Experimental results for the analysis of different axial positions of the detector. The retrieved cross-correlation images for the multiplexed PVB with TCs of $|-1\rangle+|1\rangle+|7\rangle-|8\rangle$ when the camera is placed (a) 200.0, (b) 300.0 and (c) 400.0 mm behind the RGG. The retrieved cross-correlation images for the PVB with a single TC of $|9\rangle$ when the camera is placed (d) 200.0, (e) 300.0 and (f) 400.0 mm behind the RGG.

In various application scenarios, the position variations of the detector cannot be completely avoided, leading to power loss or inter-channel crosstalk.[42] Figure 7 shows the retrieved cross-correlation images obtained by using the proposed method, when different lateral misalignments are considered. High robustness is achieved against the lateral misalignments of the detector. It is confirmed in experiments that the quality of the retrieved cross-correlation images remains high.

Figure 8 shows the retrieved cross-correlation images using the proposed method, when different axial distances of the detector are tested. It is illustrated in Fig. 8 that the pattern size in the cross-correlation images proportionally changes in consistency with the descriptions in Eq. (3). The morphological features, i.e., the number of petals and azimuthal distributions at each layer, remain stable, and different axial distances can be flexibly employed. In Fig. 3, when the larger TC $|\ell|$ is used in PVB, the ring size in MCF would be larger. However, when the axial distance between the RGG and the detector is larger, the ring size in MCF is also larger as shown in Figs. 8(d)–8(f). Therefore, conventional method[17,31] is not feasible, since the sizes need to be applied for data decoding and the axial distance between the RGG and detector needs to be fixed. Here, these limits are overcome in the proposed method, and the applicability is enhanced in real-world scenarios.

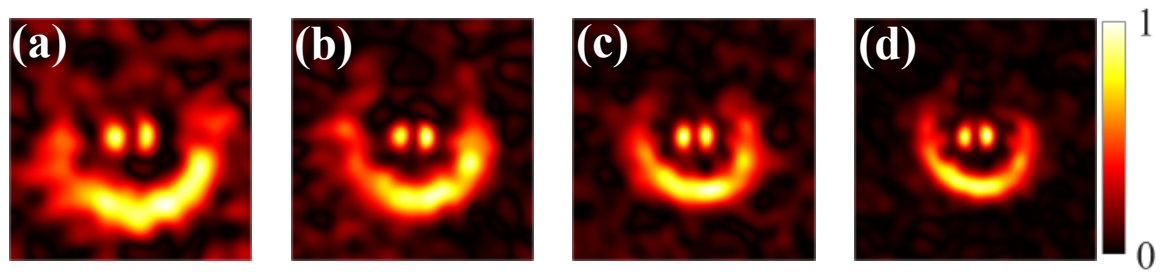

**FIG. 9.** Experimental results obtained by using multiplexed PVBs with varying sizes in the optical setup. The retrieved cross-correlation images for the multiplexed PVB with TCs of $|-1\rangle+|1\rangle+|7\rangle-|8\rangle$ using a radius $r_0$ in Eq. (1) of (a) 1.5 mm, (b) 1.8 mm, (c) 2.1 mm and (d) 2.4 mm. The retrieved cross-correlation images in (a)-(d) remain recognizable.

Figure 9 shows the retrieved cross-correlation images, when the multiplexed PVBs with different sizes are used in experiments. Here, when the radius $r_0$ of PVB is changed to 1.5 mm, $\Delta r$ in Eq. (1) is revised accordingly to be $(1.5/1.8)\times 0.83$ mm. It can be seen in Fig. 9 that as the size of the multiplexed PVB increases, the pattern size in the cross-correlation images becomes smaller. The observed phenomenon is consistent with the description in Eq. (3).

## IV. CONCLUSION

We have reported the light beam carrying OAM through time-varying scattering media using dual orthogonal-polarization channels. The multiplexed PVB serves as the signal beam, and a plane wave is used as the reference beam. 6-bit data can be encoded to be optically transmitted through time-varying scattering media, and data decoding can be implemented by analyzing the retrieved cross-correlation images. A 6-bit image with 4096 pixels is used in experiments to illustrate error-free data transmission using the proposed method, and high robustness is also achieved. The proposed method provides a promising solution for high-fidelity OAM transmission in harsh environments.

## ACKNOWLEDGMENTS

This work was supported by Hong Kong Research Grants Council (15224921, 15223522, 15237924, C5047-24G) and The Hong Kong Polytechnic University (1-CDJA, 1-WZ4M).

## AUTHOR DECLARATIONS

### Conflict of Interest

The authors have no conflicts to disclose.

### Author Contributions

**Heshen Li:** Data curation (lead); Formal analysis (lead); Investigation (lead); Methodology (lead); Writing – original draft (lead). **Jin Wei:** Formal analysis (lead); Investigation (lead). **Tianshun Zhang:** Formal analysis (lead); Investigation (lead). **Wen Chen:** Conceptualization (lead); Formal analysis (lead); Methodology (lead); Resources (lead); Funding acquisition (lead); Project administration (lead); Supervision (lead); Writing – review & editing (lead).

## DATA AVAILABILITY

The data that support the findings of this study are available from the corresponding author upon reasonable request.